\documentclass[aps,pra,twocolumn,showpacs,preprintnumbers,amsmath,amssymb,superscriptaddress]{revtex4-2}

\usepackage{graphicx}
\usepackage{epsfig}
\usepackage{amsmath}
\usepackage{amssymb}
\usepackage{dcolumn}
\usepackage{bm}
\usepackage{braket}
\usepackage{mathtools}
\usepackage[nice]{nicefrac}
\usepackage{bbm}
\usepackage[usenames,dvipsnames]{xcolor}
\usepackage[colorlinks=true,citecolor=MidnightBlue,linkcolor=MidnightBlue,urlcolor=MidnightBlue]{hyperref}
\usepackage{siunitx}

\PassOptionsToPackage{numbers,sort&compress}{natbib}


\renewcommand{\BibitemShut}[1]{} 

\DeclareMathOperator{\Tr}{Tr}

\newcommand{\op}[1]{\hat{#1}} 

\begin{document}

\newcommand*{\MAINZ}{QUANTUM, Institut f\"ur Physik, Johannes Gutenberg-Universit\"at Mainz, Staudingerweg 7, 55128 Mainz, Germany}
\newcommand*{\ERLANGEN}{Institut f\"ur Optik, Information und Photonik, Friedrich-Alexander Universit\"at Erlangen-N\"urnberg, Staudtstr. 1, 91058 Erlangen, Germany}
\newcommand*{\SAOT}{Erlangen Graduate School in Advanced Optical Technologies (SAOT), Friedrich-Alexander Universit\"at Erlangen-N\"urnberg, Paul-Gordan-Str. 6, 91052 Erlangen, Germany}
\homepage{http://www.quantenbit.de}

\title{Superradiance of non-interacting atoms}

\author{M.~Bojer}
\affiliation{Quantum Optics and Quantum Information, Institut f\"{u}r Optik, Information und Photonik, Friedrich-Alexander-Universit\"{a}t Erlangen-N\"{u}rnberg (FAU), 91058 Erlangen, Germany}
\author{J.~von~Zanthier}
\affiliation{Quantum Optics and Quantum Information, Institut f\"{u}r Optik, Information und Photonik, Friedrich-Alexander-Universit\"{a}t Erlangen-N\"{u}rnberg (FAU), 91058 Erlangen, Germany}

\date{\today}

\begin{abstract}
\noindent Fully-excited two-level atoms separated by less than the transition wavelength cooperatively emit light in a short burst, a phenomenon called superradiance by R. Dicke in 1954. The burst is characterized by a maximum intensity scaling with the square of the number of atoms $N$ and a temporal width reduced by $N$ compared to the single atom spontaneous decay time. Both effects are usually attributed to a synchronization of the electric dipole moments of the atoms occurring during the process of light emission. Contrary to this explanation, it was recently shown by use of a quantum path description that the peak intensity
results from the quantum correlations among the atoms when occupying symmetric Dicke states. 
Here we investigate from this perspective the temporal evolution of the ensemble, starting in the small sample limit, i.e., when the atoms have mutual separations much smaller than the transition wavelength $\lambda$ and pass down the ladder of symmetric Dicke states. In addition, we explore the temporal evolution for the case of non-interacting atoms with mutual separations much larger than $\lambda$.~We show that in this case a similar superradiant burst of the emitted radiation is observed if the quantum correlations of the atoms are generated by conditional photon measurements retaining the atomic ensemble within or close to the symmetric subspace.
\end{abstract}

\maketitle

\noindent A single two-level atom prepared in the excited state will exponentially decay towards the ground state and spontaneously emit a photon with a characteristic rate $\Gamma$.
Assuming not only one atom but $N$ non-interacting atoms with mutual separation much larger than the transition wavelength, i.e., with all atoms radiating independently of each other, will lead to the same exponential decay with identical decay rate as in the single atom case. This behavior changes if the atoms are placed in close proximity to each other, i.e., closer than the transition wavelength, such that they are subject to the dipole-dipole interaction. In this case, instead of an exponential decay, the light is emitted in a short burst, with a maximum intensity scaling with the square of the number of atoms $N$ and a duration decreasing linearly with $N$~\cite{gross1982}, a phenomenon referred to as superradiance by R. Dicke in 1954~\cite{Dicke1954}. After this seminal work, a plethora of theoretical investigations of superradiance and the collective emission of light of atomic ensembles have been published~\cite{Lehmberg1970II,Eberly1971,Lugiato1975,Carmichael2003,Clemens_2004,Cirac2008,Scully2010,Wiegner2011,Wiegner2015,Ruostekoski2016,Ficek2018,liberal2019grating,masson2020many}, some of them confirmed also experimentally~\cite{Skribanowitz1973,Boyd1987,Brewer1996,scheibner2007superradiance,mlynek2014observation,Laurent2015,Guerin2016,angerer2018superradiant,okaba2019superradiance}. Commonly, in the literature, the square law of the peak intensity as well as the reduced decay time is explained by a phase-locking of the electric dipole moments of the individual atoms leading to synchronization and a single macroscopic dipole building up during light emission. 
However, as already outlined by Dicke in his original publication~\cite{Dicke1954}, the atomic ensemble radiates light while passing down the ladder of symmetric Dicke states, whereby none of the Dicke states possesses an electric dipole moment. 
This fact causes the interpretation of synchronized atomic dipole moments or a macroscopic dipole moment to be highly questionable. 

In this paper, we recapitulate an alternative interpretation for the superradiant light burst, valid in the small volume limit, that does not make use of the synchronization of individual atomic dipole moments. Rather, it is based on the build-up of atomic dipole-dipole correlations present for all symmetric Dicke states, leading to a superradiant emission behavior due to quantum path interferences among the corresponding quantum states~\cite{Wiegner2011,BhattiBojer2021}. 
Moreover, we will show that a superradiant light burst can be observed also in the case of  distant non-interacting atoms if appropriate correlations among the atoms are generated by measuring photons in the far field such that the detection is unable to distinguish the individual photon source~\cite{Thiel2007,Molner2012,Bhatti2015}. 
Here again, no atomic dipole moments are present or become phase-locked,
since the atoms are far distant and do not interact.~We will investigate the role of the measurement process and show that it projects the atomic states towards the symmetric subspace each time a photon is recorded. To this aim, we will derive the trace distance to the symmetric subspace before and after the photon measurement.

The paper is organized as follows. In Sect.~\ref{sec:SSS} we review Dicke's model of superradiance in the small sample limit and apply a quantum path formalism to interpret the corresponding results. In Sect.~\ref{sec:Super_Indep_Atoms} we focus on the case of distant non-interacting atoms and show that a similar superradiant light burst as in the small sample limit can be observed if the photons scattered by the ensemble are consecutively measured in the far field. We will explain this behavior again via a quantum path formalism of the contributing quantum states. This approach is particularly apt to follow the dynamical evolution of the atomic ensemble tailored by consecutive photon measurements. We also study how the photon measurements project the atomic state towards the symmetric subspace and in this way enhance the quantum correlations. In Sect.~\ref{sec:Conclusion} we finally conclude.

\section{Small sample superradiance}
\label{sec:SSS}
\noindent In this section we recount the phenomenon of superradiance in the small sample limit as considered by Dicke in his seminal paper in 1954~\cite{Dicke1954}. 
We start by introducing the Hamiltonian of $N$ independent non-interacting two-level atoms 
\begin{equation}
	\op{H}_S = \hbar \omega_0 \sum_{\mu=1}^N \op{S}_z^{(\mu)}\,
\end{equation}
where $\op{S}_z^{(\mu)}=\frac{1}{2}(\op{S}_+^{(\mu)}\op{S}_-^{(\mu)}-\op{S}_-^{(\mu)}\op{S}_+^{(\mu)})$ is the $z$-component of the pseudo-spin operator $\op{\bm{S}}^{(\mu)}$ of the $\mu$th atom, $\mu \in \{1,...,N\}$, and $\op{S}_\pm^{(\mu)}$ are the corresponding pseudo-spin raising and lowering operators, respectively. Motivated by the Hamiltonian $\op{H}_S$ we define the collective pseudo-spin operators as
\begin{equation}
	\begin{split}
		\op{\bm{S}} &= \sum_{\mu=1}^N \op{\bm{S}}^{(\mu)}\;,\;\op{S}_z = \sum_{\mu=1}^N \op{S}_z^{(\mu)}\,,\\
		\op{S}_+ &= \sum_{\mu=1}^N \op{S}_+^{(\mu)}\;,\; \op{S}_- = \sum_{\mu=1}^N \op{S}_-^{(\mu)}\,.
	\end{split}
\end{equation}
The eigenstates of $\op{H}_S$ are characterized by the three quantum numbers $J$, $M$ and $\alpha$. In this way, we can write the eigenstates as $\ket{J,M,\alpha}$ which are also eigenstates of $\op{S}_z$ as well as $\op{S}^2$, obeying the eigenvalue equations
\begin{align}
	\op{S}_z \ket{J,M,\alpha} &= \hbar M \ket{J,M,\alpha}\,,\\
	\op{S}^2 \ket{J,M,\alpha} &= \hbar^2 J(J+1) \ket{J,M,\alpha}\,,
\end{align}
where $\alpha$ accounts for the degeneracy of a given value of $J$. The degeneracy can be shown to be~\cite{Dicke1954}
\begin{equation}
	\beta = \frac{N!(2J+1)}{(\frac{N}{2}+J+1)!(\frac{N}{2}-J)!}
\end{equation}
with $\alpha\in \lbrace 1,2,...,\beta \rbrace$, whereas the different values of $J$ and $M$ are given by
\begin{align}
	J&\in \left\lbrace 0,1,...,\frac{N}{2} \right\rbrace &&\text{for } N \text{ even}\,,\\
	J&\in \left\lbrace \frac{1}{2},\frac{3}{2},...,\frac{N}{2} \right\rbrace &&\text{for } N \text{ uneven}\,,\\
	M&\in \lbrace -J,-J+1,...,J \rbrace\,. 
\end{align}
The states $\ket{J,M,\alpha}$ are the so-called Dicke states. In the following we restrict ourselves to the symmetric subspace characterized by $J=N/2$. Here, we find $\beta=1$, so that we can omit the third index and simply write $\ket{J,M}$ ($J=N/2$).\\ 
\indent In his seminal paper Dicke considered $N$ interacting atoms confined to a volume smaller than the transition wavelength cubed. As a consequence one can neglect the phase factors of the electromagnetic field in the interaction Hamiltonian~\cite{Dicke1954,Agarwal1974}. However, we note that this symmetry property only holds for specific configurations of the atomic ensemble~\cite{gross1982}. Assuming the corresponding symmetry to be fulfilled, the total Hamiltonian including the light-matter interaction in the dipole approximation commutes with $\op{S}^2$ and $\op{S}_z$. As a consequence, the time evolution does not mix different $J$ subspaces. Correspondingly, starting with a state in the symmetric subspace, the time evolution will retain the state in the symmetric subspace at any later times.
The corresponding master equation in the interaction picture 
reads~\cite{Agarwal1974}
\begin{equation}
	\partial_t \op{\rho}=-\gamma (\op{S}_+\op{S}_-\op{\rho}+\op{\rho} \op{S}_+\op{S}_--2\op{S}_-\op{\rho} \op{S}_+)\, ,
	\label{eq:small_sample_MEQ}
\end{equation} 
where $\gamma=\Gamma/2$ is the half decay rate.  
By projecting Eq.~(\ref{eq:small_sample_MEQ}) onto the basis $\ket{J,M}$ we obtain the following rate equations for the diagonal elements of the density matrix~\cite{gross1982,Agarwal1974}
\begin{equation}
	\label{eq:Rateeq}
	\begin{split}
		\partial_t \rho_{M,M}=&-2\gamma(J+M)(J-M+1)\rho_{M,M}\\
		&+2\gamma (J+M+1)(J-M)\rho_{M+1,M+1}\,,
	\end{split}
\end{equation}
where $\rho_{M,M}=\braket{J,M|\op{\rho}|J,M}$. Note that $2\gamma(J+M)(J-M+1)$ is the rate with which the state $\ket{J,M}\bra{J,M}$ decays to the state $\ket{J,M-1}\bra{J,M-1}$ and $2\gamma (J+M+1)(J-M)$ is the rate with which it gets populated by the state $\ket{J,M+1}\bra{J,M+1}$. 

\begin{figure}
	\centering
	\includegraphics[width=.48\textwidth]{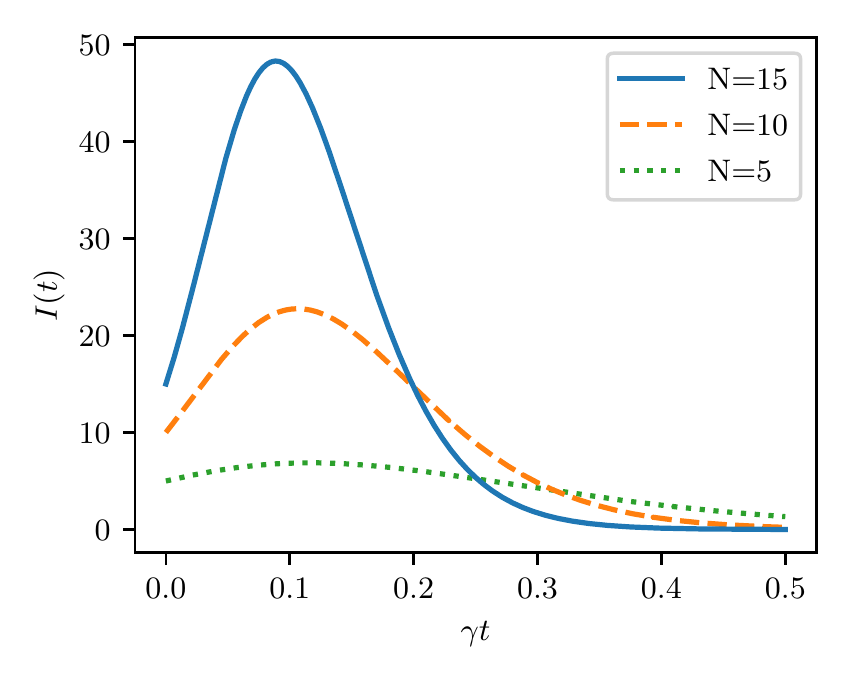}
	\caption{Intensity $I(t)$ against $\gamma t$ for an initially fully excited system $\ket{N/2,N/2}$ in case of a small sample dynamics for three different total numbers of atoms $N$. Due to the collective emission behavior one obtains a short burst instead of an exponential decay.}
	\label{fig:Small_Superradiance}
\end{figure}

Assuming the system to be initially in the fully excited state $\ket{N/2,N/2}$, the rate equations can be solved analytically. One finds that the system descends the ladder of symmetric Dicke states $\ket{N/2,M}$, with $N/2 \geq M \geq -N/2$, with a radiated intensity not following the usual exponential decay as in the case of a single atom but displaying a short burst with a peak intensity proportional to $N^2$ and a width inversely proportional to $N$ (see Fig.\,\ref{fig:Small_Superradiance}).
The usual explanation for this behavior is the build up of a macroscopic dipole moment emitting the light coherently with increasing amplitude. As can be seen from Eq.~\eqref{eq:Rateeq}, the time evolution of the atomic ensemble involves only the population of symmetric Dicke states. Since each Dicke state $\ket{J,M}$ does not carry any dipole moment, i.e., $\braket{\op{S}_\pm}=0$, this explanation appears highly questionable. In what follows, we will show that the outlined evolution of the radiated intensity can be attributed to the build-up of dipole-dipole correlations among the atoms resulting from the entanglement of the atoms in the symmetric Dicke states $\ket{J,M}$~\cite{BhattiBojer2021}. To this aim, we give a short explanation of the production of dipole-dipole correlations and the resulting intensity evolution in terms of a quantum path interpretation~\cite{Wiegner2011}.

Neglecting proportionality constants, the radiated intensity is given by~\cite{Wiegner2011}
\begin{equation}
	\begin{split}
		I =& \braket{\op{S}_+ \op{S}_-} = \sum_{\mu,\nu}\braket{\op{S}_+^{(\mu)}\op{S}_-^{(\nu)}}\\
		=&\sum_{\mu}\braket{\op{S}_+^{(\mu)}\op{S}_-^{(\mu)}}+\sum_{\substack{\mu,\nu\\ \mu\neq\nu}}\braket{\op{S}_+^{(\mu)}}\braket{\op{S}_-^{(\nu)}}\\
		&+\sum_{\substack{\mu,\nu\\ \mu\neq\nu}}\left(\braket{\op{S}_+^{(\mu)}\op{S}_-^{(\nu)}}-\braket{\op{S}_+^{(\mu)}}\braket{\op{S}_-^{(\nu)}}\right)\,,
	\end{split}
	\label{eq:Dicke_Int}
\end{equation}
where the first part describes the incoherent contribution of each atom to the intensity, the second part the coherent contribution from the non-vanishing dipole moments and the third part characterizes the quantum correlations. In the case of Dicke states Eq.~\eqref{eq:Dicke_Int} reduces to~\cite{Wiegner2011}
\begin{equation}
\label{eq:Dicke_Int_1}
	I = \sum_{\mu}\braket{\op{S}_+^{(\mu)}\op{S}_-^{(\mu)}} + \sum_{\substack{\mu,\nu\\ \mu\neq\nu}}\braket{\op{S}_+^{(\mu)}\op{S}_-^{(\nu)}}\,,
\end{equation}
since $\braket{\op{S}_\pm^{(\mu)}}=0$. This shows that there is no coherent contribution to the radiated intensity which can be assigned to a dipole moment of the atoms, but only an incoherent contribution stemming from the dipole-dipole correlations. 

As an illustration of the quantum path formalism we start by looking at the symmetric Dicke state $\ket{1,0}$ for $N=2$ atoms, i.e., $\ket{1,0}=\frac{1}{\sqrt{2}}(\ket{e,g}+\ket{g,e})$, where $\ket{e}$ and $\ket{g}$ denote the excited and ground state of the individual atoms, respectively. Plugging this state into Eq.~(\ref{eq:Dicke_Int_1}), we find
\begin{equation}
	\label{eq:Exint}
	I = 1 + 1 = 2\,.
\end{equation}
Compared to the separable states $\ket{e,g}$ or $\ket{g,e}$ alone, i.e., states with the same number of atoms in the excited state but no entanglement, the intensity of the symmetric Dicke state is enhanced. This results from the quantum correlations among the atoms due to the entanglement of the state $\ket{1,0}$~\cite{BhattiBojer2021}. More physical insight into the contribution of the individual terms is gained by a quantum path interpretation~\cite{Wiegner2011} (see Fig.~\ref{fig:QPI}):
\begin{figure}
	\centering
	\includegraphics[width=.4\textwidth]{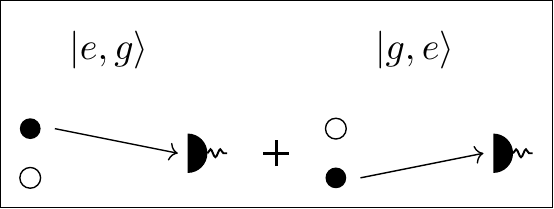}
	\caption{Quantum paths that lead to a detection event. Filled out dots correspond to atoms being in the excited state, empty dots correspond to atoms being in the ground state.}
	\label{fig:QPI}
\end{figure}
In order to record a photon, the photon must be either emitted by the first atom or the second atom leading to the two quantum paths illustrated in Fig.~\ref{fig:QPI}. After the detection of the photon has taken place the state of the atomic ensemble is the total ground state $\ket{g,g}$. Since the two quantum paths drawn in the figure lead to the same final state and are added coherently in the initial state $\ket{1,0}$, the two quantum paths  interfere with each other. Setting the electric field amplitudes to $1$ [see Eq.~(\ref{eq:Dicke_Int})], we can now find the radiated intensity: The state $\ket{1,0}$ is the sum of the two product states $\ket{e,g}$ and $\ket{g,e}$, each having a single excitation; a detection event leads only to one final state $\ket{g,g}$; therefore, the number of quantum paths per final state is 
\begin{equation}
	\label{eq:singlequpaths}
\begin{split}
	&\frac{\text{\# product states}\cdot \text{\# quantum paths per product state}}{\text{\# different final states}}\\&=\frac{2\cdot 1}{1}=2\,.
	\end{split}
\end{equation}
When calculating the total intensity, i.e., taking the modulus square of the sum of the transition amplitudes, each of the quantum paths is multiplied with itself (incoherent contribution) and in addition interferes with the other path (interference term), leading in total to $2^2$ quantum path pairs. Taking the normalization of the state $\ket{1,0}$ into account, the radiated intensity thus calculates to $(1/\sqrt{2})^2\cdot 2^2 \cdot 1=2$, where $2^2$ is the number of quantum path pairs and $1$ is the number of different final states. In this way, by just counting the number of quantum path pairs we obtain the same value as by a straightforward mathematical calculation leading to Eq.~\eqref{eq:Exint}. 

We now apply the quantum path formalism to an arbitrary symmetric Dicke state $\ket{J,M}$ and use the results obtained above to explain the phenomenon of superradiance in the small sample limit. We start by noting that $M=J-n_g=N/2-n_g$, where $n_g$ denotes the number of atoms in the ground state of the state $\ket{J,M}$. 
The intensity radiated by an atomic ensemble in the state $\ket{J,M}$ can then be calculated to
\cite{Wiegner2011}
\begin{equation}
	\label{eq:QPformula}
	I=[\mathcal{N}\cdot\mathcal{P}_{\text{no.}}^{\text{pair}}\cdot f_{\text{no.}}]_{\ket{J,M}}\,,
\end{equation}
where $\mathcal{N}$ denotes the squared normalization of the state $\ket{J,M}$, $\mathcal{P}_{\text{no.}}^{\text{pair}}$ is the number of quantum path pairs leading to the same final state and $f_{\text{no.}}$ is the number of different final states. In the case of a symmetric Dicke state $\ket{J,M}$ we find that $\mathcal{N}=\binom{N}{n_g}^{-1}$ and $f_{\text{no.}}=\binom{N}{n_g+1}$. The number of quantum path pairs $\mathcal{P}_{\text{no.}}^{\text{pair}}$ needs more explanation. The state $\ket{J,M}$ contains $\binom{N}{n_g}$ product states, where for each of these product states one has $N-n_g$ different quantum paths leading to a detection event. Thus, one finds for Eq.~\eqref{eq:singlequpaths}~\cite{Wiegner2011}
\begin{equation}
	\binom{N}{n_g}\cdot (N-n_g)/f_{\text{no.}}=\binom{N}{n_g}\cdot (N-n_g)/\binom{N}{n_g+1}=n_g+1
\end{equation} 
single quantum paths per final state. Each of these paths is multiplied with itself and in addition interferes with all the other quantum paths leading in total to $\mathcal{P}_{\text{no.}}^{\text{pair}}=(n_g+1)^2$ quantum path pairs. We thus find for the intensity~\cite{Wiegner2011}
\begin{equation}
	\begin{split}
		I&=\binom{N}{n_g}^{-1}\cdot (n_g+1)^2\cdot \binom{N}{n_g+1}\\
		&=(N-n_g)(n_g+1)=(J+M)(J-M+1) \; .
	\end{split}
\end{equation}
Note that this expression, found by counting the different interfering quantum paths leading to a detection event, gives the same result as derived from the quantum mechanical calculation $\braket{\op{S}_+\op{S}_-}=\braket{J,M|\op{S}_+\op{S}_-|J,M}=(J+M)(J-M+1)$, presented by Dicke in his seminal paper \cite{Dicke1954}. However, the quantum path interference interpretation gives a more transparent explanation apt to display the physical origin of the observed superradiant intensity.

The quantum path interpretation can be applied also to compute the time evolution of the intensity in the small sample limit. Since the master equation~\eqref{eq:small_sample_MEQ}  causes the atomic ensemble to descend the ladder of symmetric Dicke states, the time evolved state can be written as
\begin{equation} \label{eq:rho_t}
	\op{\rho}(t)=\sum_{M=-J}^J \rho_{M,M}(t)\ket{J,M}\bra{J,M}\,.
\end{equation}
The temporal evolution of the intensity can thus be expressed in mathematical terms as
\small
\begin{equation}
\label{eq.intensity-sum}
	I(t)=\Tr[\op{S}_+\op{S}_-\op{\rho}(t)] = \sum_{M=-J}^J \rho_{M,M}(t)\braket{J,M|\op{S}_+\op{S}_-|J,M}\,.
\end{equation}
\normalsize
Applying the quantum path interpretation to each state in the sum of Eq.~\eqref{eq.intensity-sum} the intensity $I(t)$ at any time $t$ can be computed. Thus, the quantum path formalism is able to explain the entire superradiant emission dynamics of the atomic ensemble in the small sample limit. In the next section we show that one can get a similar behavior in the case of distant non-interacting atoms, where the build up of quantum correlations is not caused by the time evolution of the atomic ensemble, but by consecutive photon measurements.

\section{Superradiance of far distant independent atoms}
\label{sec:Super_Indep_Atoms}
\noindent In this section we investigate the radiated intensity in the case of $N$ two-level atoms with mutual separations much larger than the transition wavelength, such that dipole-dipole interactions can be neglected. In this limit, the atomic ensemble, e.g., starting from the fully excited state, does not stay in the subspace of symmetric Dicke states. Rather, in the course of photon emissions, the system will leave the symmetric subspace  populating also non-symmetric Dicke states. However, by performing photon measurements at particular positions the atomic system can be forced back towards the symmetric subspace~\cite{Wiegner2015,Bhatti2015,BhattiBojer2021}. Measuring the photons consecutively in time the temporal evolution of the photon emission displays again a similar peaked intensity profile as in the case of the small atomic sample. The physical explanation for this behavior can once again be obtained employing the quantum path formalism. 

Note that a measurement of a photon does not create a dipole moment in the atomic ensemble. Neither does a macroscopic dipole moment build up in the course of the temporal evolution of the system due to atom-atom interactions, since the atoms are far distant from each other. The observed superradiant emission behavior thus results again only from the quantum correlations and the entanglement of the atomic ensemble, in this case generated by photon detection. 
This occurs even for atoms with separations much larger than the transition wavelength known under the name of \textit{entanglement at a distance}~\cite{moehring2007entanglement,hofmann2012,slodicka2013,bernien2013heralded,delteil2016generation,araneda2018,richter2020,richter2022,stockill2017}.

\subsection{Master equation and differential equations for the density matrix elements}
\label{sec:Meq}
\noindent Since we consider atoms with mutual separations much larger than the transition wavelength any interactions between the particles can be neglected . The master equation in the interaction picture then contains only the spontaneous decay term of the individual atoms. For $N$ atoms it reads~\cite{Agarwal1974}
\begin{equation} 
	\label{eq:indep_atoms_meq}
	\partial_t \op{\rho} = -\gamma\sum_{\mu=1}^{N}(\op{S}_+^{(\mu)}\op{S}_-^{(\mu)}\op{\rho}+\op{\rho} \op{S}_+^{(\mu)}\op{S}_-^{(\mu)}-2\op{S}_-^{(\mu)}\op{\rho} \op{S}_+^{(\mu)}) \, .
\end{equation}
Next, we project the master equation onto an orthonormal basis, in this case the bare basis given by the tensor product of the single atom states. Let $\ket{\alpha}$ and $\ket{\beta}$ denote two states of this orthonormal basis. The projection then reads (see App.~\ref{sec:AppA})
\begin{equation}
		\label{eq:dens_elem_diff_main}
	\begin{split}
		\partial_t\rho_{\alpha,\beta}(t)=&\;2\gamma\sum_{\mu=1}^{N}\rho_{\eta^{(\mu)}, \delta^{(\mu)}}(t)\\
		&-\gamma\sum_{\mu=1}^{N}\rho_{S_+^{(\mu)}\xi,\beta}(t)-\gamma\sum_{\mu=1}^N\rho_{\alpha,S_+^{(\mu)}\zeta}(t)\,,
	\end{split}
\end{equation}
where the indices of the density matrix elements on the right hand side of Eq.~\eqref{eq:dens_elem_diff_main} indicate the following conditions 
\begin{equation}
	\begin{split}
		\text{1. term:}\quad &\op{S}_-^{(\mu)}\ket{\eta}=\ket{\alpha}\,,\, \op{S}_-^{(\mu)}\ket{\delta}=\ket{\beta}\,,\\
		\text{2. term:}\quad &\op{S}_+^{(\mu)}\ket{\xi}=\ket{\alpha}\,,\\
		\text{3. term:}\quad &\op{S}_+^{(\mu)}\ket{\zeta}=\ket{\beta} \;.
	\end{split}
	\label{eq:conditions}
\end{equation}
Note that only if the conditions of Eq.~(\ref{eq:conditions}) are fulfilled a non-zero contribution from the individual terms of the right hand side of Eq.~\eqref{eq:dens_elem_diff_main} is obtained.

In what follows we 
denote the states of the bare basis of the atoms with respect to the number of atoms being in the ground state. I.e., we define the fully excited state by $\ket{e}\coloneqq\ket{e,e,...,e}$, whereas the states where atom $j$ is in the ground state and all the other atoms are in the excited state is denoted by $\ket{g_j^{(1)}}\coloneqq \op{S}_-^{(j)}\ket{e}$. States with an arbitrary number of atoms in the ground state are defined and denoted analogously.

As an example for the application of Eqs.~\eqref{eq:dens_elem_diff_main} and \eqref{eq:conditions}, we calculate the rate equation for the matrix element $\rho_{e,e}(t)$. Since it is a diagonal element, the two conditions for the first term reduce to one condition $\op{S}_-^{(\mu)}\ket{\eta}=\op{S}_-^{(\mu)}\ket{\delta}=\ket{e}$; however, since the state $\ket{e}$ is the fully excited state there exist no states $\ket{\eta}, \ket{\delta}$ which fulfill this condition, so this term vanishes. The second and third condition merge to the same condition $\op{S}_+^{(\mu)}\ket{\xi}=\op{S}_+^{(\mu)}\ket{\zeta}=\ket{e}$. Here, it is obvious that there exists only one state which fulfils this condition, namely $\ket{g_\mu^{(1)}}$. Moreover, the summation over all atoms gives a factor of $N$ for each of the two terms.
Alternatively, to follow more closely the physical process, one can interpret the equation $\op{S}_+^{(\mu)}\ket{\xi}=\op{S}_+^{(\mu)}\ket{\zeta}=\ket{e}$ with $\mu \in \lbrace 1,2,...,N \rbrace$ as a dynamical variable. In this case, one finds immediately the $N$ states which fulfill this condition, namely $\ket{g_\mu^{(1)}}$.
In both cases, the rate equation for the matrix element $\rho_{e,e}(t)$ reads
\begin{equation}
	\partial_t \rho_{e,e}(t)=-2\gamma N\rho_{e,e}(t)\,,
\end{equation}
which, since every atom decays independently, just leads to the well-known exponential decay from the state  $\rho_{e,e}(t)$ with a rate $2\gamma N$. 

Applying this formalism to all the other states characterized by the number of ground state atoms on finds a set of first-order differential equations for the density matrix elements. Assuming a number of reasonable restrictions on the initial conditions (see App.~\ref{sec:AppA}), one can reduce the number of differential equations, which can then be solved analytically to get the full consecutive time evolution of the density matrix upon conditional measurements of photons in the forward direction intersecting the free time evolutions.

\subsection{Intensity correlation functions}
\label{sec:Int_funcs}
\noindent Instead of solving the time evolution of the density matrix, let us investigate multi-photon measurements at particular locations and at consecutive times as outlined in the introduction to Sect.~\ref{sec:Super_Indep_Atoms} above. Such conditional multi-modal photon measurements are described by higher-order intensity correlation functions, where the $m$th-order intensity correlation function is defined as~\cite{Glauber1963}
\begin{widetext}
\begin{equation}
	G_{\op{\rho}}^{(m)}(\boldsymbol{r}_1,t_1,...,\boldsymbol{r}_m,t_m)=\braket{\op{E}^{(-)}(\boldsymbol{r}_1,t_1)...\op{E}^{(-)}(\boldsymbol{r}_m,t_m)\op{E}^{(+)}(\boldsymbol{r}_m,t_m)...\op{E}^{(+)}(\boldsymbol{r}_1,t_1)}_{\op{\rho}}\,.
	\label{eq:corr_func}
\end{equation}
\end{widetext}
Here, $\braket{\cdot}_{\op{\rho}}$ denotes the expectation value with respect to the quantum state $\op{\rho}$ and 
\begin{equation}
	\label{eq:Epos}
	\op{E}^{(+)}(\bm{r}_s,t_s) = \sum_{\mu=1}^{N} e^{-i k_0 \hat{\bm{r}}_s \bm{R}_\mu} \op{S}_-^{(\mu)}(t_s)
\end{equation}
is the positive frequency part of the electric field operator evaluated at position $\boldsymbol{r}_s$ and time $t_s$. Note that in Eq.~\eqref{eq:Epos}, we set the electric field amplitude to unity and used the scalar notation since we assume that the transition dipole matrix elements of the atoms are perpendicular to the detection plane. Further, $k_0$ is the wave number, $\hat{\bm{r}}_s$ is the unit vector in the direction of observation, and $\bm{R}_\mu$ and $\op{S}_-^{(\mu)}(t_s)$ are the position and the pseudo-spin lowering operator of the $\mu$th atom in the Heisenberg picture, respectively.
Since Eq.~\eqref{eq:corr_func} represents a \textit{conditional} multi-modal measurement, it can also be written in the form
\begin{widetext}
\begin{equation}
	\label{eq:Gmsplit}
	G_{\op{\rho}}^{(m)}(\boldsymbol{r}_1,t_1,...,\boldsymbol{r}_m,t_m)=G_{\op{\rho}_{m-1}}^{(1)}(\boldsymbol{r}_m,t_m)\cdot G_{\op{\rho}}^{(m-1)}(\boldsymbol{r}_1,t_1,...,\boldsymbol{r}_{m-1},t_{m-1})
\end{equation}
where~\cite{Thiel2007,Wiegner2015,carmichael2009open} 
\begin{equation}
	\op{\rho}_{m-1}=\frac{\op{E}^{(+)}(\boldsymbol{r}_{m-1},t_{m-1})...\op{E}^{(+)}(\boldsymbol{r}_1,t_1)\op{\rho} \op{E}^{(-)}(\boldsymbol{r}_1,t_1)...\op{E}^{(-)}(\boldsymbol{r}_{m-1},t_{m-1})}{\mathrm{Tr}[\op{E}^{(+)}(\boldsymbol{r}_{m-1},t_{m-1})...\op{E}^{(+)}(\boldsymbol{r}_1,t_1)\op{\rho} \op{E}^{(-)}(\boldsymbol{r}_1,t_1)...\op{E}^{(-)}(\boldsymbol{r}_{m-1},t_{m-1})]}
\end{equation}
\end{widetext}
is the state after $m-1$ photon measurements at positions $\boldsymbol{r}_1, \ldots, \boldsymbol{r}_{m-1}$ and times $t_1, \ldots, t_{m-1}$, with $\op{\rho}_0=\op{\rho}$. 
Note that $G_{\op{\rho}_{m-1}}^{(1)}$ depends on all former $m-1$ photon detection positions and detection times, since $\op{\rho}_{m-1}$ depends on all former $m-1$ photon detection events.

In what follows, we assume for simplicity that the atoms are arranged along the $x$-axis and that all detectors are located in the far field at the same position on the $y$-axis. In addition, we suppose the atomic dipole moments pointing towards  the $z$-axis, i.e., perpendicular to the detection plane ($xy$-plane). The intensity correlation functions then only depend on the different detection times $t_1, \ldots, t_{m-1}$, but not on the particular detection positions $\boldsymbol{r}_1, \ldots, \boldsymbol{r}_{m-1}$, since the positive electric field operator in the far field in this case is simply given by $\op{E}^{(+)}(\boldsymbol{r}_s,t_s)= \op{S}_-(t_s)$ [see Eq.~\eqref{eq:Epos}]. In this way, neither geometric phases nor atomic dipole patterns need to be accounted for in the calculation of $G_{\op{\rho}}^{(m)}(\boldsymbol{r}_1,t_1,...,\boldsymbol{r}_m,t_m)$. As a consequence, we can calculate simply the multi-time correlations of the collective pseudo-spin raising and lowering operators. This allows in particular to compare the result with the small sample case discussed in Sect.~\ref{sec:SSS}.

Rewriting Eq.~\eqref{eq:Gmsplit} as
\begin{equation}
	\label{eq:CondG}
	G_{\op{\rho}_{m-1}}^{(1)}(t_m)=\frac{G_{\op{\rho}}^{(m)}(t_1,...,t_m)}{G_{\op{\rho}}^{(m-1)}(t_1,...,t_{m-1})}\,,
\end{equation} 
where we have already neglected the position dependency, and applying Eq.~\eqref{eq:spincorrQRT} of App.~\ref{sec:AppB} involving the quantum regression theorem~\cite{Agarwal1974,carmichael2009open} leads to
\begin{equation}
	\label{eq:corrsub}
	\begin{split}
		&G_{\rho_{m-1}}^{(1)}(t_m)=\frac{\prod_{i=1}^{m}e^{-2\gamma t_i} \braket{\overbrace{\op{S}_+(0)...\op{S}_+(0)}^{m\text{ times}}\overbrace{\op{S}_-(0)...\op{S}_+(0)}^{m\text{ times}}}}{\prod_{i=1}^{m-1}e^{-2\gamma t_i} \braket{\underbrace{\op{S}_+(0)...\op{S}_+(0)}_{m-1\text{ times}}\underbrace{\op{S}_-(0)...\op{S}_+(0)}_{m-1\text{ times}}}}\\
		&= e^{-2\gamma t_m}\braket{\op{S}_+(0)\op{S}_-(0)}_{\op{\rho}_{m-1}}=e^{-2\gamma t_m}G_{\op{\rho}_{m-1}}^{(1)}(0)\,.
	\end{split}
\end{equation}
Therefore, to compute $G_{\rho_{m-1}}^{(1)}(t_m)$ we only have to calculate the correlation function $G_{\op{\rho}_{m-1}}^{(1)}(0)$ at time $t=0$. 

\begin{figure}
	\centering
	\includegraphics[width=.48\textwidth]{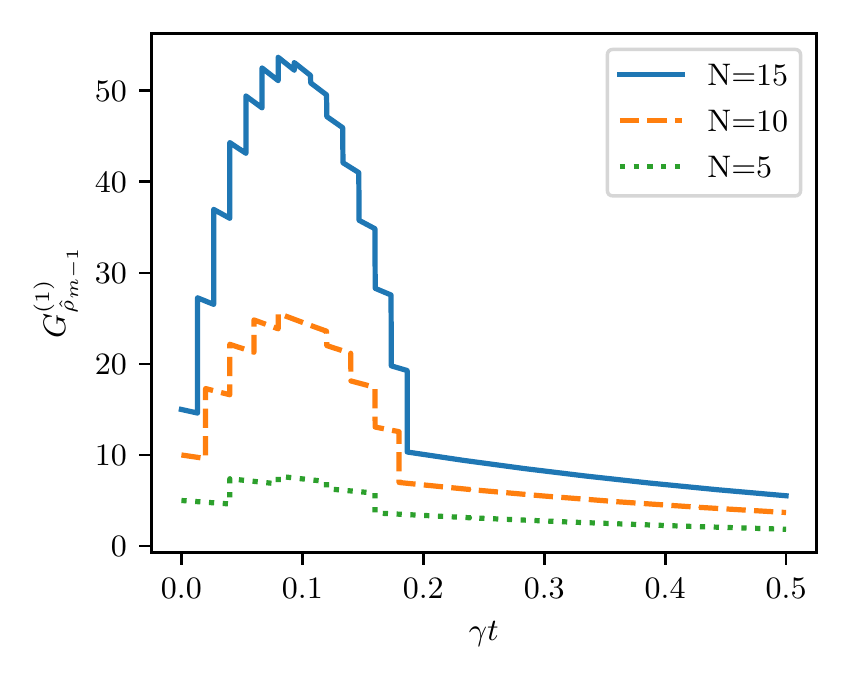}
	\caption{Conditional intensities $G_{\rho_{m-1}}^{(1)}(t_m)$ against $\gamma t$ for an initially fully excited system $\ket{e}$ with $m=1$ to $m=N$ in consecutive steps for three different $N$.}
	\label{fig:Superradiance_dist_atoms}
\end{figure}

Assuming an initially fully excited system, the correlation function  at time $t=0$ is simple to derive. This is due to the fact that for $\hat{\rho}=\ket{e}\bra{e}$ the state after $m$ photon detections at $t=0$ is given by $\hat{\rho}_{m}=\ket{J,M}\bra{J,M}$, where $J=N/2$ and $M=N/2-m$, i.e., the measurement causes the system to descent down the ladder of symmetric Dicke states. In this way, we obtain for  $G_{\op{\rho}_{m-1}}^{(1)}(0)$
\begin{equation}
	\begin{split}
		&\braket{\op{S}_+(0)\op{S}_-(0)}_{\op{\rho}_{m-1}}=\braket{J,M|\op{S}_+\op{S}_-|J,M}\\
		&=(J+M)(J-M+1)=m\cdot (N-m+1)\,.
	\end{split}
	\label{eq:coeff}
\end{equation}
In Fig.~\ref{fig:Superradiance_dist_atoms}, the conditional intensities $G_{\op{\rho}_{m-1}}^{(1)}(t_m)$ are plotted against $\gamma t$ for the initially fully excited system. The different intersections of the conditional intensity correspond to the consecutive times at which a spontaneously emitted photon is detected.
We see that the measurement process leads to steps in the conditional intensity, resulting from the quantum correlations among the atoms produced each time a photon is recorded. For instance, for $N=5$, the five consecutive conditional intensities $G_{\op{\rho}_{m-1}}^{(1)}(t_m)$ are given by [see Eq.~(\ref{eq:coeff})]
\begin{equation}
	\begin{split}
		G_{\op{\rho}_{0}}^{(1)}(t_1) &= 5\cdot e^{-2\gamma t_1}\,,\\
		G_{\op{\rho}_{1}}^{(1)}(t_2) &= 8\cdot e^{-2\gamma t_2}\,,\\
		G_{\op{\rho}_{2}}^{(1)}(t_3) &= 9\cdot e^{-2\gamma t_3}\,,\\
		G_{\op{\rho}_{3}}^{(1)}(t_4) &= 8\cdot e^{-2\gamma t_4}\,,\\
		G_{\op{\rho}_{4}}^{(1)}(t_5) &= 5\cdot e^{-2\gamma t_5}\,.
	\end{split}
	\label{eq:subsections}
\end{equation} 
From Eq.~(\ref{eq:subsections}), we see that after every intersection the intensity follows the usual single-atom decay at a rate $2\gamma$ 
resulting from the fact that the atoms are non-interacting (see Fig.~\ref{fig:Superradiance_dist_atoms}). 
However, due to the increase of the corresponding coefficients, as calculated in Eq.~(\ref{eq:coeff}),
a superradiant burst like the one in the small sample limit is observed, even for distant atoms, if conditional photon measurements are considered. In the next section we show that this superradiant behavior can be explained again by the quantum path formalism. 

\subsection{Quantum path interference interpretation}
\label{sec:quantum_path}
\noindent Starting from the differential equations for the density matrix elements in the case of distant atoms [see Eq.~\eqref{eq:dens_elem_diff_main} and App.~\ref{sec:AppA}] we can analytically solve the time evolution of the density operator and in this way compute the temporal evolution of the intensity in the forward direction as considered in Sect.~\ref{sec:Int_funcs}. Alternatively, the latter can be derived employing the quantum path formalism. This allows to give a transparent physical explanation of the time-dependent intensity, in particular if conditional measurements are involved.

We assume again that the initial state is the fully excited state $\ket{e}$. For this state we obviously have $\mathcal{N}=1$, $\mathcal{P}_{\text{no.}}^{\text{pair}}=1$ and $f_{\text{no.}}=N$, so that the initial coefficient equals $N$ [see Eq.~\eqref{eq:QPformula}].
Further, from the differential equations for the density matrix elements we obtain for the density operator $\op{\rho}_0(t_1)$ at time $t_1$
\begin{widetext}
\begin{equation}
	\op{\rho}_0(t_1)=p_e(t_1)\ket{e}\bra{e}+p_{g^{(1)}}(t_1)\sum_j \ket{g_j^{(1)}}\bra{g_j^{(1)}}+p_{g^{(2)}}(t_1)\sum_{j<k}\ket{g_{jk}^{(2)}}\bra{g_{jk}^{(2)}}+\dots\,,
	\label{eq:evol_1}
\end{equation}
\end{widetext}
where the probabilities, obtained by solving the differential equations in App.~\ref{sec:AppA}, are given by 
\begin{equation}
	\begin{split}
		p_e(t_1)&=e^{-2N\gamma t_1}\\
		p_{g^{(1)}}(t_1)&=e^{-2N\gamma t_1}\cdot(e^{2\gamma t_1}-1)\\
		p_{g^{(2)}}(t_1)&=e^{-2N\gamma t_1}\cdot(e^{2\gamma t_1}-1)^2\\
		&\phantom{h}\vdots\\
		p_{g^{(s)}}(t_1)&=e^{-2N\gamma t_1}\cdot (e^{2\gamma t_1}-1)^s\,,
	\end{split}
	\label{eq:probab}
\end{equation}
with $\ket{g^{(0)}}=\ket{e}$.
We note that due to the particular initial conditions (see App.~\ref{sec:AppA}) and the photon measurement in forward direction all states with the same number of ground state atoms evolve with the same time behavior; in this case, we can write the probabilities listed in Eq.~\eqref{eq:probab} simply in front of the individual sums appearing in Eq.~\eqref{eq:evol_1}.
 
Next, we derive the intensity radiated in the forward direction at time $t_1$ using in Eq.~\eqref{eq:evol_1} the quantum path interference interpretation for each term with the same number of ground state atoms. For each of these terms we have $\mathcal{N}=1$. Thus, we only need to compute for each of these terms the number of interfering quantum path pairs and the number of final states. For a single product state with $n_g$ atoms in the ground state we have $N-n_g$ excited states, i.e., $N-n_g$ possible single quantum paths. Further, for each product state we have exactly $\binom{N-n_g}{1}$ different final states, so that $\mathcal{P}_{\text{no.}}^{\text{pair}}=1^2$. Consequently, the coefficient in front of each block of the density matrix with the same number of ground state atoms becomes according to Eq.~\eqref{eq:QPformula}
\begin{equation}
	\binom{N}{n_g}\cdot \mathcal{N}\cdot\mathcal{P}_{\text{no.}}^{\text{pair}}\cdot f_{\text{no.}} = \binom{N}{n_g}\cdot 1 \cdot 1^2 \cdot \binom{N-n_g}{1}\,,
\end{equation}
where $\binom{N}{n_g}$ is the number of terms in each block with the same number of ground state atoms.
The correlation function for the first measurement thus reads
\small
\begin{align}
		G_{\op{\rho}_{0}}^{(1)}(t_1)&=\sum_{s=0}^{N-1} e^{-2N\gamma t_1}\cdot (e^{2\gamma t_1}-1)^s \cdot \binom{N}{s}\cdot 1 \cdot 1^2 \cdot \binom{N-s}{1}\nonumber\\
		&= 1\cdot (N-1+1) e^{-2\gamma t_1}=N e^{-2\gamma t_1}\,,
			\label{eq:inten}
\end{align}
\normalsize
where we wrote the second-last expression in Eq.~\eqref{eq:inten} as in Eq.~\eqref{eq:coeff}.

To find the state after the first measurement, we have to apply the collective spin lowering and raising operators onto the time-evolved state Eq.~\eqref{eq:evol_1}. The individual states in Eq.~\eqref{eq:evol_1} then transform to
\small
\begin{equation}
	\begin{split}
		\ket{g^{(0)}}&\rightarrow \ket{g^{(0)}_{S_-}}\coloneqq S_-\ket{g^{(0)}}=\sum_{j=1}^N S_-^{(j)}\ket{e}= \sum_{j=1}^N \ket{g_j^{(1)}}\\
		\ket{g_j^{(1)}}&\rightarrow \ket{g_{j,S_-}^{(1)}}\coloneqq S_-\ket{g_j^{(1)}}=\sum_{k=1}^N S_-^{(k)}\ket{g_j^{(1)}}=\sum_{\substack{k=1\\ k\neq j}}^N \ket{g_{jk}^{(2)}}\\
		\ket{g_{jk}^{(2)}}&\rightarrow \ket{g_{jk,S_-}^{(2)}}\coloneqq S_-\ket{g_{jk}^{(2)}}=\sum_{l=1}^N S_-^{(l)}\ket{g_{jk}^{(2)}}=\sum_{\substack{l=1\\ l\neq j, l \neq k}}^N\ket{g_{jkl}^{(3)}}\\
		&\phantom{h}\vdots
	\end{split}
\end{equation}
\normalsize
Solving the corresponding differential equations we can write the time-evolved state after the first photon measurement as
\begin{widetext}
\begin{equation}
\label{eq:photon-subtracted state}
		\op{\rho}_1(t_2)=\frac{1}{N}\left\lbrace p_{g^{(0)}_{S_-}}(t_2) \ket{g^{(0)}_{S_-}}\bra{g^{(0)}_{S_-}}+ p_{g_{S_-}^{(1)}}(t_2) \sum_{j} \ket{g_{j,S_-}^{(1)}}\bra{g_{j,S_-}^{(1)}}+p_{g_{S_-}^{(2)}}(t_2)\sum_{j,k} \ket{g_{jk,S_-}^{(2)}}\bra{g_{jk,S_-}^{(2)}}+\dots\right\rbrace\,,
\end{equation}
\end{widetext}
where the probabilities are independent of the index and given by
\begin{equation}
	p_{g^{(s)}_{S_-}}(t_2) = e^{-2(N-1)\gamma t_2}(e^{2\gamma t_2}-1)^s\,.
\end{equation} 
Using Eq.~\eqref{eq:photon-subtracted state}, we can again apply the quantum path interference formalism to calculate the intensity after measurement of the first photon. The normalization is $\mathcal{N}=1/N$. Furthermore, a state originating from a previous state with $N-n_g+1$ excitations is the sum of $N-n_g+1$ different states giving rise to interference. Note that here $n_g$ denotes the number of atoms being in the ground state after the first measurement, not the initial number of ground state atoms, i.e., $1\leq n_g\leq N-1$. Since we consider the first-order intensity correlation function after the measurement of the first photon, i.e., the conditional second-order correlation function [see Eq.~\eqref{eq:CondG}] we have to distribute two ground states among all initially excited atoms, leading to $\binom{N-n_g+1}{2}$ different final states. Again, we have $N-n_g$ quantum paths for each different product state with $n_g$ atoms in the ground state (see above).~Thus, we find $\mathcal{P}_{\text{no.}}^{\text{pair}}=2^2=4$ interfering quantum path pairs so that the coefficient of each block reads
\begin{equation}
	\binom{N}{n_g-1}\cdot \frac{1}{N} \cdot 2^2 \cdot \binom{N-n_g+1}{2}\,.
\end{equation}
Finally, to calculate the conditional second-order intensity correlation function, i.e., the first-order intensity correlation function after the measurement of the first photon, we have to sum over all contributions in Eq.~\eqref{eq:photon-subtracted state}. Shifting the summation index down by one, we find
\begin{widetext}
\begin{equation}
\label{eq:inten1}
		G_{\op{\rho}_1}^{(1)}(t_2)=\sum_{s=0}^{N-2} e^{-2(N-1)\gamma t_2}\cdot (e^{2\gamma t_2}-1)^s \cdot  \binom{N}{s} \cdot \frac{1}{N} \cdot 2^2 \cdot \binom{N-s}{2}= 2(N-2+1) e^{-2\gamma t_2} = 2(N-1)e^{-2\gamma t_2}\,,
\end{equation}
\end{widetext}
where again we wrote the second-last expression in Eq.~\eqref{eq:inten1} as in Eq.~\eqref{eq:coeff}. The presented method to find the conditional intensity correlation function after measurement of the first photon can be applied to an arbitrary order $m$. In this case the general probabilities are found to be
\begin{equation}
	p_s(t_m)=e^{-2(N-m+1)\gamma t_m}(e^{2\gamma t_m}-1)^s\,,
	\label{eq:probgen}
\end{equation}
where we characterize the probability by the initial number of ground state atoms $s$ of the time-evolved state Eq.~\eqref{eq:evol_1} before the first measurement. The normalization for an arbitrary correlation order $m$ is given by
\begin{equation}
	\mathcal{N}=\frac{1}{[(m-1)!]^2}\cdot \binom{N}{m-1}^{-1}\,.
\end{equation}
Considering subsequent measurement processes, we find for states with $N-n_g$ excitations $\binom{N-n_g+(m-1)}{m}$ different final states, where the number of ground state atoms $n_g$ refers to the number of ground state atoms before the last measurement. The number of interfering terms can then be counted to
\begin{equation}
\underbrace{(m-1)!}_{\text{multiple counting}}\underbrace{\binom{N-n_g+(m-1)}{m-1}}_{\text{interfering quantum states}}\,,
	\label{eq:intterms}
\end{equation}
where we split up the expression into a factor that accounts for the multiple counting of the same quantum state and the number of different interfering quantum states. Note that the multiple counting factor squared cancels exactly with a factor coming from the normalization. The number of single quantum paths per final state thus calculates to 
\begin{equation}
	\frac{\binom{N-n_g+(m-1)}{m-1} \cdot (N-n_g)}{\binom{N-n_g+(m-1)}{m}}=m\,,
\end{equation}
so that the number of interfering quantum path pairs is given by $m^2$. The $m$th-order intensity correlation function can thus finally be written as
\begin{widetext}
\begin{equation}
	\label{eq:intens-m}
			G_{\op{\rho}_{m-1}}^{(1)}(t_m)=\sum_{s=0}^{N-m} e^{-2(N-m+1)\gamma t_m}(e^{2\gamma t_m}-1)^s \binom{N}{s}\cdot \binom{N}{m-1}^{-1} \cdot m^2 \cdot \binom{N-s}{m} = m(N-m+1)\cdot e^{-2\gamma t_m}\,,
\end{equation}
\end{widetext}
which is the same result as the one  calculated by use of the quantum regression theorem [see Eqs.~\eqref{eq:corrsub} and \eqref{eq:coeff} above]. However, in the derivation of Eq.~\eqref{eq:intens-m}, we explicitly solved the first-order differential equations for the density matrix elements and then applied the quantum path formalism, what allows for a transparent physical interpretation. In this way, we are able to explain the occurring superradiant behavior of distant atoms if consecutive measurements of the radiated photons are considered. In the next section we analyze the role of the measurement process more quantitatively and show that it projects the time-evolved states towards the symmetric superradiant subspace. The latter implies the appearance of quantum correlations leading to the observed coherent emission of spontaneous radiation.

\begin{figure}
	\centering
	\includegraphics[width=.48\textwidth]{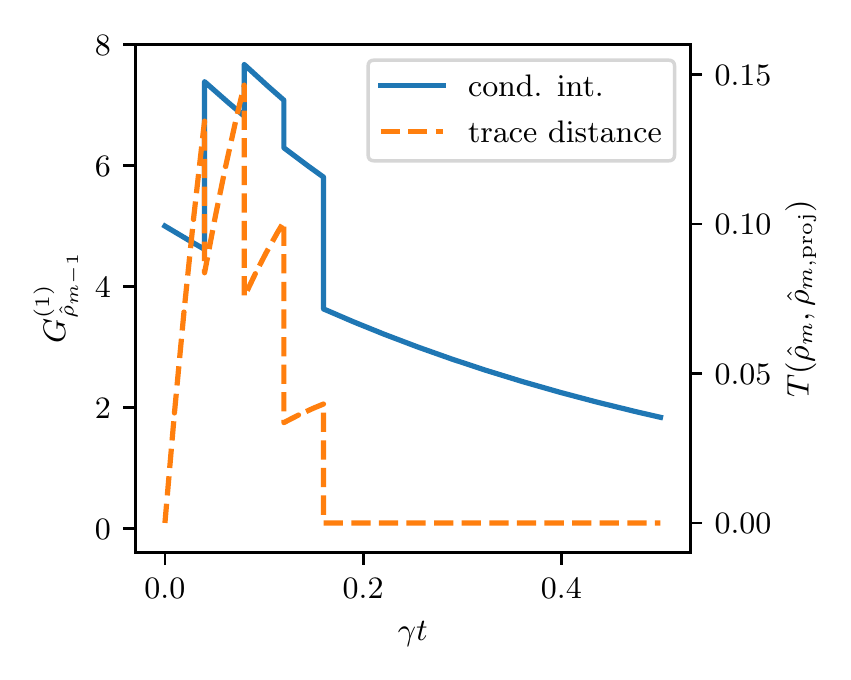}
	\caption{Conditional intensities $G_{\rho_{m-1}}^{(1)}(t_m)$ against $\gamma t$ for an initially fully excited system $\ket{e}$ for $N=5$ and $m=1$ to $m=5$ in consecutive steps (left axis) together with the trace distance Eq.~\eqref{eq:Trdistfinal} (right axis). The measurement projects the state closer to the symmetric subspace leading to a similar course of the conditional intensities as in the small sample case.}
	\label{fig:Trace_Dist_5}
\end{figure}

\subsection{Role of measurement process - distance to symmetric subspace}
\label{sec:Dist_sym}
\noindent In this subsection we investigate the distance of the time-evolved conditioned states obtained after consecutive measurement of $m$ photons to the symmetric subspace spanned by the symmetric Dicke states $\ket{J,M}$, with $J=N/2$ and $M=N/2-n_g$. To that aim, let us calculate the trace distance 
\begin{equation}
	T(\hat{\rho}_{m},\hat{\rho}_{m,\text{proj}})=\frac{1}{2}\Tr\left[\sqrt{(\hat{\rho}_{m}-\hat{\rho}_{m,\text{proj}})^\dagger (\hat{\rho}_{m}-\hat{\rho}_{m,\text{proj}})}\right]
	\label{eq:Tracedist}
\end{equation}
between the time-evolve state $\hat{\rho}_{m}$ obtained after the measurement of $m$ photons and its projection onto the symmetric subspace
\begin{equation}
	\hat{\rho}_{m,\text{proj}}=\sum_{M,M'}\ket{J,M}\braket{J,M|\hat{\rho}_{m}|J,M'}\bra{J,M'}\,.
	\label{eq:rhoproj0}
\end{equation}
In this way we obtain the final result for the trace distance (see App.~\ref{sec:AppC})
\begin{widetext}
\begin{equation}
	T(\hat{\rho}_{m},\hat{\rho}_{m,\text{proj}})=\frac{1}{2}\sum_{n_g=0}^{N}\binom{N}{n_g}\text{Time}(m,n_g)[\text{Diag}(m,n_g)-\text{Diag}_{\text{proj}}(m,n_g)]\,,
	\label{eq:Trdistfinal}
\end{equation}
\end{widetext}
where the different functions are defined in App.~\ref{sec:AppC}. In Figure~\ref{fig:Trace_Dist_5} we show for $N=5$ the trace distance between the time-evolved states after the measurement of $m$ photons, $m \in \{1,...,N\}$, and the corresponding states projected onto the symmetric subspace, together with the conditional intensities Eq.~\eqref{eq:intens-m}. We see that each photon measurement projects the state closer to the symmetric subspace resulting in a temporal behavior of the conditional intensities similar to the one observed in the small sample case.

\section{Conclusion}
\label{sec:Conclusion}
\noindent In conclusion, we have shown that Dicke superradiance consists of two different aspects. One is the magnitude of the radiated intensity, one is the duration of the radiated pulse. Starting from the small sample limit, we explicitly showed that the magnitude of the radiated intensity does not arise from a synchronization of the individual dipole moments of the atoms, as the expectation value of the dipole moment is zero for all Dicke states. It rather stems from the quantum correlations of the atoms which build up in the course of the dynamic evolution of the system. The superradiant behavior, i.e., the light burst and its peaked intensity, can eventually be attributed to the interference of the indistinguishable quantum paths resulting from the entanglement of the symmetric Dicke states. The shortened time period of the light burst can be explained by the altered transition rates [see Eqs.~\eqref{eq:Rateeq} and \eqref{eq:rho_t}].

In a next step, we applied the quantum path interference explanation to the case of far distant non-interacting  atoms. While quantum correlations naturally arise via the coupled dynamics of the atoms in the small sample case, in the configuration of distant atoms they can be generated by consecutive conditional photon measurements. The latter lead again to entanglement between the individual atoms and thus to  quantum coherences. These coherences then give rise to interference effects which can again be explained using the quantum path interference formalism. The superradiant burst of the radiated intensity can thus be adequately described also in the case of distant non-interacting atoms by the interference of indistinguishable quantum paths. However, in contrast to the small sample case, the period of the superradiant burst obtained via conditional measurements depends uniquely on the respective measurement times. 

\section*{Acknowledgements}
\noindent M.B. and J.v.Z. gratefully acknowledge funding and support by the International Max Planck Research School - Physics of Light. This work was funded by the Deutsche Forschungsgemeinschaft (DFG, German Research Foundation) -- Project-ID 429529648 -- TRR 306 QuCoLiMa ("Quantum Cooperativity of Light and Matter'').

\appendix
\section{Differential equations for the density matrix elements}
\label{sec:AppA}
\noindent If we project the master equation Eq.~\eqref{eq:indep_atoms_meq} onto the orthonormal basis given by the tensor product states, we find
\begin{widetext}
\begin{equation}
	\label{eq:dens_elem_diff}
	\begin{split}
		\partial_t \braket{\alpha|\op{\rho}|\beta}=&-\gamma\sum_{\mu=1}^N (\braket{\alpha|\op{S}_+^{(\mu)}\op{S}_-^{(\mu)}\op{\rho}|\beta}+\braket{\alpha|\op{\rho} \op{S}_+^{(\mu)}\op{S}_-^{(\mu)}|\beta}-2\braket{\alpha|\op{S}_-^{(\mu)}\op{\rho} \op{S}_+^{(\mu)}|\beta})\\
		=&-\gamma\sum_{\mu=1}^N \left(\sum_{\xi}\braket{\alpha|\op{S}_+^{(\mu)}\ket{\xi}\bra{\xi}\op{S}_-^{(\mu)}\op{\rho}|\beta}+\sum_{\zeta}\braket{\alpha|\op{\rho} \op{S}_+^{(\mu)}\ket{\zeta}\bra{\zeta}\op{S}_-^{(\mu)}|\beta}-2 \sum_{\eta,\delta}\braket{\alpha|\op{S}_-^{(\mu)}\ket{\eta}\bra{\eta}\op{\rho}\ket{\delta}\bra{\delta} \op{S}_+^{(\mu)}|\beta}\right)\\
		\partial_t\rho_{\alpha,\beta}(t) \stackrel{\mathclap{(\ast)}}{=}&\;	2\gamma\sum_{\mu=1}^{N}\rho_{\eta^{(\mu)}, \delta^{(\mu)}}(t)-\gamma\sum_{\mu=1}^{N}\rho_{S_+^{(\mu)}\xi,\beta}(t)-\gamma\sum_{\mu=1}^N\rho_{\alpha,S_+^{(\mu)}\zeta}(t)\,,
	\end{split}
\end{equation}
\end{widetext}
where $\ket{\alpha}$ and $\ket{\beta}$ are two tensor product states and the equality $(\ast)$ is restricted by the different conditions of Eq.~\eqref{eq:conditions} that have to be fulfilled to get a non-zero contribution from the corresponding terms on the right hand side of the last line of Eq.~\eqref{eq:dens_elem_diff}.\\
\indent In the following, we restrict ourselves to initial density matrices which do not have coherences between states with different number of ground state atoms, i.e., we restrict ourselves to block-diagonal density matrices with respect to the number of atoms being in the ground state. Moreover, we only consider initial density matrices, which provide for density matrix elements following the same kind of differential equation the same initial conditions.~We note that by starting with an initial state that fulfills these two conditions, also the states after time evolution and consecutive photon measurements in forward direction fulfill these conditions. This allows us to reduce the number of real first-order differential equations from $2^{2N}-1$ to $N^2$, i.e., we only need to find the time evolution of particular elements of the blocks of the density matrix.\\
\indent Using as basis the tensor product basis, a straightforward calculation of the differential equations of the density matrix elements leads to the following first-order differential equation for the diagonal elements
\begin{widetext}
\begin{equation}
	\partial_t
	\begin{pmatrix}
		\rho_{e,e}(t)\\
		\rho_{g_j^{(1)},g_j^{(1)}}(t)\\
		\rho_{g_{jk}^{(2)},g_{jk}^{(2)}}(t)\\
		\vdots
	\end{pmatrix}
	=
	\begin{pmatrix}
		-2\gamma N & 0 & 0 & \dots\\
		2\gamma & -2\gamma(N-1) & 0 &\\
		0 & 4\gamma & -2\gamma(N-2) &\\
		\vdots & & & \ddots
	\end{pmatrix}
	\begin{pmatrix}
		\rho_{e,e}(t)\\
		\rho_{g_j^{(1)},g_j^{(1)}}(t)\\
		\rho_{g_{jk}^{(2)},g_{jk}^{(2)}}(t)\\
		\vdots
	\end{pmatrix}\,.
	\label{eq:matrix_rate_diag}
\end{equation}
\end{widetext}
In an analogous procedure one can find the differential equation for the off-diagonal elements, which is given by
\small
\begin{widetext}
	\begin{equation}
		\partial_t
		\begin{pmatrix}
			\rho_{g_j^{(1)},g_k^{(1)}}(t)\\
			\rho_{g_{jk}^{(2)},g_{jl}^{(2)}}(t)\\
			\rho_{g_{jk}^{(2)},g_{lm}^{(2)}}(t)\\
			\rho_{g_{jkl}^{(3)},g_{jkm}^{(3)}}(t)\\
			\rho_{g_{jkl}^{(3)},g_{jmn}^{(3)}}(t)\\
			\rho_{g_{jkl}^{(3)},g_{mno}^{(3)}}(t)\\
			\vdots
		\end{pmatrix}
		=
		\begin{pmatrix}
			-2\gamma (N-1) & 0 & 0 & 0 & 0 & 0 & \dots\\
			2\gamma & -2\gamma(N-2) & 0 & 0 & 0 & 0 &\\
			0 & 0 & -2\gamma(N-2) & 0 & 0 & 0 &\\
			0 & 4\gamma & 0 & -2\gamma(N-3) & 0 & 0 &\\
			0 & 0 & 2\gamma & 0 & -2\gamma(N-3) & 0 &\\
			0 & 0 & 0 & 0 & 0 & -2\gamma(N-3) &\\
			\vdots & & & & & & \ddots
		\end{pmatrix}
		\begin{pmatrix}
			\rho_{g_j^{(1)},g_k^{(1)}}(t)\\
			\rho_{g_{jk}^{(2)},g_{jl}^{(2)}}(t)\\
			\rho_{g_{jk}^{(2)},g_{lm}^{(2)}}(t)\\
			\rho_{g_{jkl}^{(3)},g_{jkm}^{(3)}}(t)\\
			\rho_{g_{jkl}^{(3)},g_{jmn}^{(3)}}(t)\\
			\rho_{g_{jkl}^{(3)},g_{mno}^{(3)}}(t)\\
			\vdots
		\end{pmatrix}\,.
		\label{eq:matrix_rate_off}
	\end{equation}
\end{widetext}
\normalsize
Solving this reduced number of first-order differential equations by either taking the matrix exponential or a successive application of the Duhamel-principle allows us to fully determine the time evolution of an ensemble of independent atoms for the above described initial states analytically.

\section{Multi-time correlations of raising and lowering pseudo-spin operators}
\label{sec:AppB}
\noindent We first investigate the $m$th-order single-time correlation function $\braket{\op{S}_+\op{S}_+...\op{S}_+\op{S}_-...\op{S}_-}_{\op{\rho}(t)}$ in a state $\op{\rho}(t)$ that is block-diagonal with respect to the number of atoms in the ground state, i.e., each block has a well-defined number of ground state atoms. Writing the expectation value explicitly
\begin{widetext}
\begin{equation}
	\braket{\op{S}_+\op{S}_+...\op{S}_+\op{S}_-...\op{S}_-}_{\op{\rho}(t)} = \sum_{i_1=1}^N...\sum_{i_m=1}^N \sum_{j_1=1}^N...\sum_{j_m=1}^N \braket{\op{S}_+^{(i_1)}...\op{S}_+^{(i_m)}\op{S}_-^{(j_1)}...\op{S}_-^{(j_m)}}_{\op{\rho}(t)}
\end{equation}
\end{widetext}
leads us to the evaluation of expectation values of the form $\braket{\op{S}_+^{(i_1)}...\op{S}_+^{(i_m)}\op{S}_-^{(j_1)}...\op{S}_-^{(j_m)}}_{\op{\rho}(t)}$. If either $i_s=i_p$ or $j_s=j_p$ for some $s$ and $p$ the expectation value vanishes, since $[\op{S}_-^{(\mu)}]^2=[\op{S}_+^{(\mu)}]^2=0$ $\forall \mu \in \lbrace 1,...,N \rbrace$. In addition, if any $i_s\neq j_p\;\forall p$ or $j_p\neq i_s\;\forall s$ the expectation  value also vanishes, since we assume a block-diagonal state with respect to the number of ground state atoms implying that there are no coherences between states with different number of atoms in the ground state. Therefore, we can restrict ourselves to expectation values of the form $\left\langle\prod_{s=1}^m \op{S}_+^{(i_s)} \prod_{p=1}^m \op{S}_-^{(i_p)}\right\rangle_{\rho(t)}$ with $i_s\neq i_p\;\forall s\neq p$. To calculate these $m$th-order correlation functions we first investigate the simple case of the first-order expectation value $\braket{\op{S}_+^{(i)}\op{S}_-^{(i)}}_{\op{\rho}(t)}$. We calculate the time derivative
\begin{widetext}
\begin{equation}
	\begin{split}
		\frac{d}{d t} \braket{\op{S}_+^{(i)}\op{S}_-^{(i)}}_{\op{\rho}(t)} =&\,\, \frac{d}{d t} \Tr[\op{S}_+^{(i)}\op{S}_-^{(i)}\op{\rho}(t)]=\Tr[\op{S}_+^{(i)}\op{S}_-^{(i)}\dot{\op{\rho}}(t)]\\
		=&-\gamma\sum_{\mu=1}^N\left\lbrace \Tr[\op{S}_+^{(i)}\op{S}_-^{(i)}\op{S}_+^{(\mu)}\op{S}_-^{(\mu)}\op{\rho}(t)]+\Tr[\op{S}_+^{(i)}\op{S}_-^{(i)}\op{\rho}(t) \op{S}_+^{(\mu)}\op{S}_-^{(\mu)}]-2\Tr[\op{S}_+^{(i)}\op{S}_-^{(i)}\op{S}_-^{(\mu)}\op{\rho}(t) \op{S}_+^{(\mu)}] \right\rbrace\,,
	\end{split}
\end{equation}
\end{widetext}
where we used the master equation Eq.~\eqref{eq:indep_atoms_meq}. To find the differential equation the expectation value fulfils we can do the following simplifications
\begin{widetext}
\begin{align}
	&\begin{rcases}
		\Tr[\op{S}_+^{(i)}\op{S}_-^{(i)}\op{S}_+^{(\mu)}\op{S}_-^{(\mu)}\op{\rho}(t)]=\Tr[\op{S}_+^{(i)}\op{S}_-^{(i)}\op{\rho}(t) \op{S}_+^{(\mu)}\op{S}_-^{(\mu)}]\;\hphantom{=\Tr[\op{S}_+^{(i)}\op{S}_-^{(i)}\op{\rho}(t)]}\\
		\Tr[\op{S}_+^{(i)}\op{S}_-^{(i)}\op{S}_-^{(\mu)}\op{\rho}(t) \op{S}_+^{(\mu)}]=\Tr[\op{S}_+^{(i)}\op{S}_-^{(i)}\op{\rho}(t) \op{S}_+^{(\mu)}\op{S}_-^{(\mu)}]
	\end{rcases}
	\text{for}\; i\neq \mu\,,\\
	&\begin{rcases}
		\Tr[\op{S}_+^{(i)}\op{S}_-^{(i)}\op{S}_+^{(\mu)}\op{S}_-^{(\mu)}\op{\rho}(t)]=\Tr[\op{S}_+^{(i)}\op{S}_-^{(i)}\op{\rho}(t) \op{S}_+^{(\mu)}\op{S}_-^{(\mu)}]=\Tr[\op{S}_+^{(i)}\op{S}_-^{(i)}\op{\rho}(t)]\\
		\Tr[\op{S}_+^{(i)}\op{S}_-^{(i)}\op{S}_-^{(\mu)}\op{\rho}(t) \op{S}_+^{(\mu)}]=0
	\end{rcases}
	\text{for}\; i= \mu\,,
\end{align}
\end{widetext}
such that for $i\neq \mu$ all terms in the sum cancel each other, whereas for $i=\mu$ we get $2\braket{\op{S}_+^{(i)}\op{S}_-^{(i)}}_{\op{\rho}(t)}$. Thus, we find the differential equation
\begin{equation}
	\frac{d}{d t} \braket{\op{S}_+^{(i)}\op{S}_-^{(i)}}_{\op{\rho}(t)} = -2\gamma \braket{\op{S}_+^{(i)}\op{S}_-^{(i)}}_{\op{\rho}(t)}\,,
\end{equation}
which is solved by $\braket{\op{S}_+^{(i)}\op{S}_-^{(i)}}_{\op{\rho}(t)}=e^{-2\gamma t}\braket{\op{S}_+^{(i)}\op{S}_-^{(i)}}_{\op{\rho}(0)}$. Now we can generalize this result to get the higher-order equal-time correlation functions. Therefore, we first consider the second-order correlation function $\braket{\op{S}_+^{(i_1)}\op{S}_+^{(i_2)}\op{S}_-^{(i_2)}\op{S}_-^{(i_1)}}_{\op{\rho}(t)}$. Since $i_1\neq i_2$ we have the same situation as before, with the difference that now we get two contributions from $i_1=\mu$ and $i_2=\mu$, such that we immediately find
\small
\begin{equation}
	\frac{d}{dt} \braket{\op{S}_+^{(i_1)}\op{S}_+^{(i_2)}\op{S}_-^{(i_2)}\op{S}_-^{(i_1)}}_{\op{\rho}(t)} =-4\gamma \braket{\op{S}_+^{(i_1)}\op{S}_+^{(i_2)}\op{S}_-^{(i_2)}\op{S}_-^{(i_1)}}_{\op{\rho}(t)}\,.
\end{equation}
\normalsize
In general, we get~\cite{thiel2009quantum} \small
\begin{align}
	\frac{d}{dt} \left\langle\prod_{s=1}^m \op{S}_+^{(i_s)} \prod_{p=1}^m \op{S}_-^{(i_p)}\right\rangle_{\op{\rho}(t)} = -2m\gamma \left\langle\prod_{s=1}^m \op{S}_+^{(i_s)} \prod_{p=1}^m \op{S}_-^{(i_p)}\right\rangle_{\op{\rho}(t)}
\end{align}
\normalsize for which the solution is \small
\begin{align}
	\left\langle\prod_{s=1}^m \op{S}_+^{(i_s)} \prod_{p=1}^m \op{S}_-^{(i_p)}\right\rangle_{\op{\rho}(t)}=e^{-2m\gamma t}\left\langle\prod_{s=1}^m \op{S}_+^{(i_s)} \prod_{p=1}^m \op{S}_-^{(i_p)}\right\rangle_{\op{\rho}(0)}\,.
\end{align}
\normalsize
With this result we can now calculate the multi-time correlation functions by use of the quantum regression theorem, which relates the multi-time expectation values to single-time expectation values~\cite{Agarwal1974,carmichael2009open}. We find,
\begin{widetext}
\begin{equation}
		\frac{d}{d\tau}\braket{\op{S}_+^{(i_1)}(t)\op{S}_+^{(i_2)}(t+\tau) \op{S}_-^{(i_2)}(t+\tau)\op{S}_-^{(i_1)}(t)}=-2\gamma \braket{\op{S}_+^{(i_1)}(t)\op{S}_+^{(i_2)}(t+\tau) \op{S}_-^{(i_2)}(t+\tau)\op{S}_-^{(i_1)}(t)}\,,
\end{equation}
such that 
\begin{equation}
	\braket{\op{S}_+^{(i_1)}(t_1)\op{S}_+^{(i_2)}(t_2) \op{S}_-^{(i_2)}(t_2)\op{S}_-^{(i_1)}(t_1)}=e^{-2\gamma(t_2-t_1)}\braket{\op{S}_+^{(i_1)}(t_1)\op{S}_+^{(i_2)}(t_1) \op{S}_-^{(i_2)}(t_1)\op{S}_-^{(i_1)}(t_1)}
\end{equation}
\end{widetext} 
with $t=t_1$ and $\tau=t_2-t_1$ and we are left with a single-time expectation value, which we solved beforehand. A successive application of the quantum regression theorem to the $m$th-order multi-time correlation functions then gives~\cite{thiel2009quantum}
{\small
\begin{widetext}
\begin{equation}
	\begin{split}
		\label{eq:spincorrQRT}
		\braket{\op{S}_+^{(i_1)}(t_1)...\op{S}_+^{(i_m)}(t_m)\op{S}_-^{(i_m)}(t_m)...\op{S}_-^{(i_1)}(t_1)}&=e^{-2\gamma(t_m-t_{m-1})}e^{-4\gamma(t_{m-1}-t_{m-2})}...e^{-2m\gamma t_1}\cdot \braket{\op{S}_+^{(i_1)}(0)...\op{S}_+^{(i_m)}(0)\op{S}_-^{(i_m)}(0)...\op{S}_-^{(i_1)}(0)}\\
		&=\prod_{l=1}^m e^{-2\gamma t_l} \braket{\op{S}_+^{(i_1)}(0)...\op{S}_+^{(i_m)}(0)\op{S}_-^{(i_m)}(0)...\op{S}_-^{(i_1)}(0)}\,.
	\end{split}
\end{equation}
\end{widetext}
}

\section{Trace distance between post-measurement states and symmetric subspace}
\label{sec:AppC}
\noindent In this Appendix, we explicitly calculate the trace distance between the time-dependent post-measurement states and the symmetric subspace. Therefore, we note that due to the restrictions on the initial state the post-measurement states $\hat{\rho}_{m}$ are block-diagonal with respect to the number of ground state atoms, i.e., we can write
\begin{equation}
	\hat{\rho}_{m}= \bigoplus\limits_{n_g}\hat{\rho}_{m,n_g}\,.
\end{equation}
Since $M,M'$ correspond to a specific number of ground state atoms via $M=N/2-n_g$ the states $\ket{J,M},\ket{J,M'}$ only act on the subspace spanned by the tensor product states with the same number of ground state atoms. Thus, Eq.~\eqref{eq:rhoproj0} reduces to
\begin{widetext}
\begin{equation}
	\hat{\rho}_{m,\text{proj}}=\sum_{n_g}\braket{J,M|\hat{\rho}_{m,n_g}|J,M}\ket{J,M}\bra{J,M}=\bigoplus\limits_{n_g}\braket{J,M|\hat{\rho}_{m,n_g}|J,M}\ket{J,M}\bra{J,M}\,.
	\label{eq:rhoproj1}
\end{equation}
\end{widetext}
In addition, again due to the restrictions on the initial state the sum of the elements of a given row of $\hat{\rho}_{m,n_g}$ gives the same value for each row. Therefore, the states $\ket{J,M}$ with corresponding number of ground state atoms $n_g$ are eigenstates of the different block matrices $\hat{\rho}_{m,n_g}$. Thus, we can write
\begin{equation}
	\hat{\rho}_{m,n_g} = \braket{J,M|\hat{\rho}_{m,n_g}|J,M}\ket{J,M}\bra{J,M}\oplus \hat{\tilde{\rho}}_{m,n_g}\,.
\end{equation}
The post-measurement states can then be written as
\begin{equation}
	\hat{\rho}_{m}=\bigoplus\limits_{n_g}  \braket{J,M|\hat{\rho}_{m,n_g}|J,M}\ket{J,M}\bra{J,M}\oplus \hat{\tilde{\rho}}_{m,n_g}
\end{equation}
and using Eq.~\eqref{eq:rhoproj1} 
\begin{align}
	\label{eq:diff_mat}
	\hat{\tilde{\rho}}_{m}\coloneqq \hat{\rho}_{m}-\hat{\rho}_{m,\text{proj}} = \bigoplus\limits_{n_g} \hat{\tilde{\rho}}_{m,n_g}\,.
\end{align}
Since $\hat{\rho}_{m}$ and $\hat{\rho}_{m,\text{proj}}$ are Hermitian $\hat{\tilde{\rho}}$ is also Hermitian and we can rewrite Eq.~\eqref{eq:Tracedist} to
\begin{equation}
	T(\hat{\rho}_{m},\hat{\rho}_{m,\text{proj}})=\frac{1}{2}\Tr\left[\sqrt{\hat{\tilde{\rho}}_{m}^2}\right]\,. 
\end{equation}
Furthermore, $\hat{\tilde{\rho}}$ can be diagonalized via some unitary matrix $U$, such that
\begin{equation}
	\hat{\tilde{\rho}}_{m,D} = U^\dagger \hat{\tilde{\rho}}_{m} U
\end{equation}
is diagonal. Then,
\begin{equation}
	\sqrt{\hat{\tilde{\rho}}_{m}^2} = U \sqrt{\hat{\tilde{\rho}}_{m,D}^2} U^\dagger
\end{equation} 
and
\small
\begin{equation}
	\Tr\left[\sqrt{\hat{\tilde{\rho}}_{m}^2}\right] = \Tr\left[U \sqrt{\hat{\tilde{\rho}}_{m,D}^2} U^\dagger\right]=\Tr\left[ \sqrt{\hat{\tilde{\rho}}_{m,D}^2} \right]=\sum_{i}|\lambda_{m,i}|\,,
\end{equation}
\normalsize
where $\lambda_{m,i}$ are the eigenvalues of $\hat{\tilde{\rho}}_{m}$.\\
\indent Since $\hat{\rho}_{m}$ is a density matrix, all eigenvalues of $\hat{\rho}_{m}$ are greater than or equal to zero. Thus, also all eigenvalues of the block-matrices $\hat{\rho}_{m,n_g}$ and $\hat{\tilde{\rho}}_{m,n_g}$ are greater than or equal to zero, such that the same holds true for $\hat{\tilde{\rho}}_{m}$. Then, we find
\begin{equation}
	\begin{split}
	T(\hat{\rho}_{m},\hat{\rho}_{m,\text{proj}})&=\frac{1}{2}\Tr\left[\sqrt{\hat{\tilde{\rho}}_{m}^2}\right]=\frac{1}{2}\sum_{i}|\lambda_{m,i}|\\
	&=\frac{1}{2}\sum_{i}\lambda_{m,i}=\frac{1}{2}\Tr[\hat{\tilde{\rho}}_{m}]\,,
	\end{split}
\end{equation}
such that we only need to compute the trace of the difference matrix Eq.~\eqref{eq:diff_mat}. Therefore, we separately identify the diagonal entries of $\hat{\rho}_{m}$ and $\hat{\rho}_{m,\text{proj}}$ combinatorially. Both have the same time dependence coming from the evolution of the density matrix via the master equation. We define this factor as
\begin{equation}
	\text{Time}(m,n_g)\coloneqq e^{-2(N-m)\gamma t_{m+1}}(e^{2\gamma t_{m+1}}-1)^{n_g-m}\,.
	\label{eq:Time}
\end{equation}
Note that the difference with respect to Eq.~\eqref{eq:probgen} is that $m=1$ corresponds already to the first post-measurement state, whereas in Eq.~\eqref{eq:probgen} $m=1$ corresponds to the initial time-evolved state without a measurement done. In addition, $s$ in Eq.~\eqref{eq:probgen} denotes the initial number of ground state atoms. Here, $n_g$ denotes the current number of ground state atoms, such that a subtraction of $m$ is needed. Now we are left with counting the occurrence of diagonal states. In Eq.~\eqref{eq:intterms} we have already identified a multiple counting factor of quantum states coming from the successive application of measurements. Therefore, we find
\begin{equation}
	\binom{n_g}{n_g-m}(m!)^2 = \binom{n_g}{m}(m!)^2
\end{equation}
many occurrences of the same diagonal entry, where $\binom{n_g}{n_g-m}$ accounts for the number of initial states leading to the considered final diagonal state and $m!$ comes from the multiple counting. Note that the square of $m!$ enters, since one gets $m!$ from the "ket"-vector and $m!$ from the "bra"-vector. Accounting for the post-measurement normalization we identify the prefactor of the diagonal entries of $\hat{\rho}_{m}$ to be
\begin{equation}
	\text{Diag}(m,n_g)\coloneqq \frac{\binom{n_g}{m}(m!)^2}{\binom{N}{m}(m!)^2}=\frac{\binom{n_g}{m}}{\binom{N}{m}}\,.
\end{equation}
For $\hat{\rho}_{m,\text{proj}}$ the diagonal entries are given by $\braket{J,M|\hat{\rho}_{m,n_g}|J,M}\cdot \mathcal{N}$, where $\mathcal{N}=\binom{N}{n_g}^{-1}$ is the squared normalization of the state $\ket{J,M}$. For the scalar product we find
\begin{widetext}
\begin{equation}
	\braket{J,M|\hat{\rho}_{m,n_g}|J,M} = (\#\text{inc. terms})\cdot p_{\text{single}} \cdot (\#\text{coh. terms})^2 \cdot \text{Time}(m,n_g) \cdot \frac{1}{(m!)^2}\binom{N}{m}^{-1}\,,
\end{equation}
\end{widetext}
where $\text{Time}(m,n_g)$ denotes the time evolution factor Eq.~\eqref{eq:Time} and $\frac{1}{(m!)^2}\binom{N}{m}^{-1}$ is the post-measurement normalization. In addition, the number of initial incoherent terms is
\begin{equation}
	(\#\text{inc. terms}) = \binom{N}{n_g-m}
\end{equation}
and the number of coherent terms is given by Eq.~\eqref{eq:intterms}
\begin{equation}
	(\#\text{coh. terms}) = \prod_{i=1}^m [N-(n_g-i)]=m!\binom{N-n_g+m}{m}\,.
\end{equation}
Note that the square accounts again for the "ket" as well as the "bra" contribution. Last, $p_{\text{single}}$ denotes the probability of finding a single tensor product state within the state $\ket{J,M}$, which can easily be computed to
\begin{equation}
	p_{\text{single}}=\mathcal{N}=\binom{N}{n_g}^{-1}\,.
\end{equation}
With the above considerations we define
\footnotesize
\begin{equation}
	\text{Diag}_{\text{proj}}(m,n_g)\coloneqq \binom{N}{n_g-m}\binom{N}{n_g}^{-2}\binom{N-n_g+m}{m}^2 \binom{N}{m}^{-1}\,,
\end{equation}
\normalsize
which gives the prefactor of the diagonal entries for the projected state $\hat{\rho}_{m,\text{proj}}$.~Now, putting everything together we arrive at the final result for the trace distance
\begin{widetext}
\begin{equation}
	T(\hat{\rho}_{m},\hat{\rho}_{m,\text{proj}})=\frac{1}{2}\sum_{n_g=0}^{N}\binom{N}{n_g}\text{Time}(m,n_g)[\text{Diag}(m,n_g)-\text{Diag}_{\text{proj}}(m,n_g)]\,.
\end{equation}
\end{widetext} 


\bibliography{library}

\end{document}